\documentclass[a4paper,12pt]{article}
\usepackage{epsfig}
\def\1{\'{\i}}
\begin{document}

\title{\bf Approximating the Amplitude and Form of Limit Cycles
in the Weakly Nonlinear Regime of Li\'{e}nard Systems}

\author{J.L. L\'{o}pez$^{\dag}$ and R. L\'{o}pez-Ruiz$^{\ddag}$ \\
                                  \\
{$^{\dag}$\small Department of Mathematics and Informatics,} \\
{\small Universidad P\'ublica de Navarra, 31006-Pamplona (Spain).} \\
{$^{\ddag}$\small Department of Computer Science and BIFI,} \\
{\small Universidad de Zaragoza, 50009-Zaragoza (Spain).}
\date{ }}

\small
\maketitle
\baselineskip 8mm
 \begin{center} {\bf Abstract} \end{center}
Li\'{e}nard equations, $\ddot{x}+\epsilon f(x)\dot{x}+x=0$, with $f(x)$ an even 
continuous function are considered. In the weakly nonlinear regime ($\epsilon\rightarrow 0$), 
the number and a ${\mathcal O}(\epsilon^0)$  approximation of the
amplitude of limit cycles present in this type of systems, can be obtained by applying a methodology recently
proposed by the authors [L\'opez-Ruiz R, L\'opez JL. Bifurcation curves of limit cycles in some Li\'enard
systems. Int J Bifurcat Chaos 2000; 10:971-980]. In the present work, that method is carried forward to higher
orders in $\epsilon$ and is embedded in a general recursive algorithm capable to approximate the form of the
limit cycles and to correct their amplitudes as an expansion in powers of $\epsilon$. Several examples showing
the application of this scheme are given.

$\;$\newline
{\small {\bf Keywords:} Li\'{e}nard equation, limit cycles, weakly nonlinearity}.\newline
{\small {\bf PACS numbers:} 05.45.-a, 02.30.Hq, 02.30.Mv} \newline
{\small {\bf AMS Classification:} 37E99, 37C27, 37C50}

\newpage

\section{Introduction}

Self-sustained periodic oscillations are common in nature \cite{winfree}.
They are also important in engineering applications \cite{andronov}.
The calculation of the number and amplitude of such different periodic motions
(limit cycles) taking place in an oscillating system is an unsolved problem.
This question constitutes the second part of
Hilbert's Sixteenth Problem \cite{ilyashenko,li} when we are restricted
to two-dimensional autonomous systems of the form:
\begin{eqnarray}
 \dot{x} & = & P_n(x,y), \nonumber \\
 \dot{y} & = & Q_n(x,y),
 \label{eq1}
\end{eqnarray}
where $\dot{x}(t)=dx(t)/dt$, $\dot{y}(t)=dy(t)/dt$ and
$(P_n, Q_n)$ are polynomials of degree $n$ with real
coefficients. Although it has been proved that the number
of limit cycles in systems of type (\ref{eq1}) is finite \cite{ecalle,ilya},
the determination of the maximal number $H(n)$ of such solutions for a given
degree $n$ is far away of being known. Even for $n=2$, $H(2)$ is still not
determined \cite{ye}. It has been verified with different examples,
for instance, that $H(2)\geq 4$, $H(3)\geq 11$,
$H(4)\geq 15$ and $H(5)\geq 23$ but the exact values of
$H(n)$ for these cases \cite{ye,li1,zhang,li2} are unknown.

The van der Pol oscillator $\ddot{x}+\epsilon (x^2-1)\dot{x}+x=0$,
is an example of system (\ref{eq1}) that has been well studied.
In this case, $P_3(x,y)=y$ and $Q_3(x,y)=-\epsilon (x^2-1)y-x$.
It displays a limit cycle whose uniqueness and non-algebraicity
has been shown for the whole range of the parameter $\epsilon$ \cite{odani}.
Its behaviour runs from near-harmonic oscillations for $\epsilon$
close to zero $(\epsilon\rightarrow 0)$ to relaxation oscillations when
$\epsilon$ tends to infinity $(\epsilon\rightarrow\infty)$, making
it a good model for many practical situations
\cite{lopezruiz}.

A generalization of the van der Pol oscillator
is the Li\'{e}nard equation,
\begin{equation}
 \ddot{x}+\epsilon f(x)\dot{x}+x=0,
 \label{eq2}
\end{equation}
with $\epsilon$ a real parameter and $f(x)$ any real function.
When $f(x)$ is a polynomial
of degree $N=2n+1$ or $2n$ this equation takes the form (1)
with $P_{N+1}(x,y)=y$ and $Q_{N+1}(x,y)=-\epsilon f(x)y-x$.
It has been conjectured by Lins, Melo and Pugh (LMP-conjecture)
that the maximum number of limit cycles allowed is just $n$ \cite{lins}.
It is true if $N=2$, or $N=3$
or if $f(x)$ is even and $N=4$ \cite{lins,rychkov}.
Also, there are strong arguments for claiming its truth 
in the strongly nonlinear regime
$(\epsilon\rightarrow\infty)$ when $f(x)$ is an even
polynomial \cite{lopez}. There are no general results
about the limit cycles when $f(x)$ is a polynomial
of degree greater than $5$ neither, in general, when $f(x)$
is an arbitrary real function \cite{giacomini}.

The calculation of the form of limit cycles
is another difficult task associated to this problem.
As far as we know, there is no a general methodology
documented in the literature on this subject.
The different schemes used to calculate the number and
amplitude of the periodic motions in the weak nonlinear regime,
namely the averaging and perturbation methods \cite{andronov,jordan},
could also be applied to find the approximate wave form of those orbits
in the time or frequency domain \cite{padin}.
In general, the insistence in performing these calculations in phase space
with the time variable being explicit can complicate the achievement
of the objective.

In this work, we undertake the task of calculating
the form of limit cycles in phase space for the weakly nonlinear regime
of Li\'enard equations. The method exploited in \cite{lopezruiz1,lopez1}
to find the number and a first order approximation to the amplitude of limit 
cycles when $\mid\epsilon\mid\ll 1$
is recalled in Section 2. The embedding of this method in a general recursive
framework for calculating the form and amplitude of those periodic solutions
is presented in Section 3. The general algorithm is rewritten in Section 4.
Some illustrative examples are given in Section 5.
Finally, we present our conclusions.

\section{Integral equation for the limit cycles}

In order to study the limit cycles of equation (\ref{eq2}) with the time
variable being implicit,
it is convenient to rewrite it in the coordinates $(x,\dot{x})=(x,y)$ in the plane, 
with $\dot{x}(t)=y(x)$ and $\ddot{x}(t)=y(x)y'(x)$ (where $y'(x)=dy/dx$):
\begin{equation}
 yy'+\epsilon f(x)y+x = 0.
 \label{eq3}
\end{equation}
A limit cycle $C_l\equiv (x,y_{\pm}(x))$ of equation (\ref{eq3})
has a positive branch $y_+(x)>0$ and a negative branch $y_-(x)<0$.
They cut the $x$-axis in two points $(-a_-,0)$ and $(a_+,0)$
with $a_-,a_+>0$ because
the origin $(0,0)$ is the only fixed point of Eq. (\ref{eq3}).
Then every limit cycle $C_l$ solution of Eq. (\ref{eq3}) encloses
the origin and the oscillation $x$ runs in the interval
$-a_-<x<a_+$. The amplitudes of oscillation $a_-,a_+$ identify
the limit cycle. The result is a nested set of closed curves
that defines the qualitative distribution
of the integral curves in the plane $(x,y)$. The stability
of the limit cycles is alternate. For a given stable limit cycle,
the two neighbouring  limit cycles, the closest one in its interior
and the closest one in its exterior, are unstable,
and viceversa.

When $f(x)$ is an even function, the {\it inversion symmetry}
$(x,y)\leftrightarrow (-x,-y)$ of Eq. (\ref{eq3}) implies
$y_+(x)=-y_-(-x)$ and then $a_1=a_2=a$.
Therefore we can restrict ourselves to the positive branches
of the limit cycles $(x,y_+(x))$ with $-a\leq x \leq a$. In this case,
the amplitude $a$ identifies the limit cycle.
The {\it parameter inversion symmetry}
$(\epsilon,x,y)\leftrightarrow (-\epsilon,x,-y)$
implies that if $C_l\equiv (x,y_{\pm}(x))$ is a limit
cycle for a given $\epsilon$, then
$\overline{C}_l\equiv (x,-y_{\mp}(x))$ is a limit cycle
for $-\epsilon$. In consequence, the amplitude $a$ of the limit cycles
in these Li\'enard systems is an even function of $\epsilon$.
Moreover if $C_l$ is stable (or unstable)
then $\overline{C}_l$ is unstable
(or stable, respectively). Therefore it is enough
to consider the limit
cycles when $\epsilon>0$ for obtaining all the periodic solutions.
(The limit cycles for a given
$\epsilon <0$ are obtained from a reflection over
the $x$-axis of those limit cycles obtained for $\epsilon>0$).

Another global property of a limit cycle of Eq. (\ref{eq3})
can be derived from the fact that the mechanical energy $E=(x^2+y^2)/2$
is conserved in a half oscillation:
\begin{equation}
\int_{-a}^a\frac{dE}{dx} dx = 0.
\end{equation}
By Eq. (\ref{eq3}), we have $\frac{dE}{dx}=-\epsilon f(x)y$
and, hence, by substituting it in the last expression,
an integral equation is obtained for the limit cycle, that is:
\begin{equation}
\int_{-a}^{a}f(x)y_+(x) dx =0.
\label{eq4}
\end{equation}
The finite set of limit cycles of Eq. (\ref{eq3}) also verifies
Eq. (\ref{eq4}). If the limit cycles are expanded
in a power series of $\epsilon$, Eq. (\ref{eq3}) imposes the
differential relationships that must be verified for the different
orders in $\epsilon$; this is exploited in Sections 3.2 and 3.3. 
An alternative point of view comes from the necessary integral 
condition ($\ref{eq4}$). It can also
be exploited to find the correct approximate amplitudes
of the limit cycles up to a given order in $\epsilon$; this is carried 
out in Section 3.1 up to the order ${\mathcal O}(\epsilon^0)$ and in Section 4 up to any order.

\section{Amplitude and form of the limit cycles}

We proceed now to explain how to calculate recursively the form and amplitude of
limit cycles of Eq. (\ref{eq3}) for different orders in $\epsilon$ when
this parameter is small and $f(x)$ is even.

\subsection{Limit cycles at order zero}

For a given $y(x)$ we define the function $S(a)$ as follows,
\begin{equation}
\label{cicle1}
 S(a)\equiv\int_{-a}^af(x)y(x)dx.
\end{equation}
Then, as it has been established in Eq. (\ref{eq4}) of the previous section,
a necessary condition for $y_+(x)$, hereinafter called $y(x)$,
to be a limit cycle of Eq. (\ref{eq3}) with amplitude $a$ is
\begin{equation}
\label{cicle2}
 S(a) = 0.
\end{equation}

When $\epsilon \to 0$, the limit cycle solutions of Eq. (\ref{eq3})
emerging from the period annulus surrounding the center $(0,0)$
become circles on the plane,
\begin{equation}
y_0(x)=\sqrt{a^2-x^2}.
\end{equation}
The amplitudes of these limit cycles, $a$, verify $S(a)=0$:
\begin{equation}
\label{beta}
 \beta_0(a)\equiv\int_{-a}^af(x)y_0(x)dx=\int_{-a}^af(x)\sqrt{a^2-x^2}dx=0,
\end{equation}
and every solution $a$ of $\beta_0(a)=0$ is the germ of at least one limit cycle
of amplitude $a$ \cite{lopezruiz1}. This is a well established result if we
recall at this point \cite{li} that $\beta_0(a)$ coincides with the
first Melnikov function or, equivalently, with the Abelian integral defined for the
perturbed Hamiltonian system (\ref{eq2}) whose level curves at $\epsilon=0$ are
given by $x^2+y^2=a^2$. For instance,
when $f(x)$ is an even polynomial of degree $2n$, $\beta_0(a)$ is $a^2$ times
a polynomial of degree $n$ in $a^2$.
Then, it has at most $n$ simple positive roots, and we can conclude that
LMP-conjecture is true in this regime \cite{lopezruiz1}.

\subsection{Limit cycles up to order ${\mathcal O}(\epsilon^N)$}

By continuity, we can assume that the limit cycle $y(x)$ for small $\epsilon$
is approximated by $y_0(x)$ plus an expansion in powers of $\epsilon$:
\begin{equation}
\label{expan}
 y(x)=y_0(x)+\sum_{n=1}^{N} \epsilon^ny_n(x)+{\mathcal O}(\epsilon^{N+1}).
\end{equation}
Replacing this expansion in (\ref{eq3}) and equating powers of $\epsilon$,
every function $y_n(x)$ of the expansion satisfies
a first order linear differential equation of the form:
\begin{equation}
\label{yene}
(y_0y_n)'+W_{n-1}=0, \hskip 1cm n=1,2,3,...,N,
\end{equation}
where
\begin{equation}
W_{n-1}(x)\equiv \left\lbrace
\begin{array}{ll}
y_0(x)f(x) & {\rm if}  \hskip 2mm n=1, \\
\sum_{k=1}^{n-1} y'_k(x)y_{n-k}(x)+y_{n-1}(x)f(x) &
{\rm if} \hskip 2mm 2\le n \le N.
\end{array}
\right.
\end{equation}
 The system of equations (\ref{yene}) 
can be iteratively solved for $y_n(x)$ by imposing the contour condition $y_n(-a)=0$ $\forall$ $n$.
Then, we obtain:
\begin{eqnarray}
y_n(x) & = & -{1\over y_0(x)}\int_{-a}^xW_{n-1}(t)dt = \nonumber \\
 & = &  -{1\over y_0(x)}\left\lbrace {1\over
2}\sum_{k=1}^{n-1}y_k(x)y_{n-k}(x)+\int_{-a}^xy_{n-1}(t)f(t)dt\right\rbrace.
\label{eqw}
\end{eqnarray}
Therefore, for small $\epsilon$, the solutions $y(x)$ of Eq. (\ref{eq3}), which
verify $y(-a)=0$, are approximated up to the order ${\mathcal O}(\epsilon^N)$ by
\begin{equation}
y(x)\simeq y^{(N)}(x)\equiv \sum_{n=0}^{N} \epsilon^ny_n(x).
\label{eqyn}
\end{equation}

For example, up to the order ${\mathcal O}(\epsilon^3)$, $y(x)$ can be written as
\begin{equation}
y^{(3)}(x)= y_0(x)+\epsilon y_1(x)+\epsilon^2 y_2(x)+\epsilon^3 y_3(x),
\label{y3}
\end{equation}
where
\begin{eqnarray}
\label{yonee}
y_1(x) & = & -{1\over y_0(x)}\int_{-a}^xy_0(t)f(t)dt, \\
\label{ytwo}
 y_2(x) & = & -{y_1^2(x)\over 2y_0(x)}-{1\over y_0(x)}\int_{-a}^xy_1(t)f(t)dt, \\
\label{ythree}
 y_3(x) & = &-{y_1(x)y_2(x)\over y_0(x)}-{1\over y_0(x)}\int_{-a}^xy_2(t)f(t)dt.
\end{eqnarray}

The possible amplitudes $a$ of the limit cycles of Eq. (\ref{eq3}),
up to the order ${\mathcal O}(\epsilon^{N})$,
are  determined
by imposing that the solution $y^{(N)}(x)$ vanishes also at $x=a$:
\begin{equation}
\label{conditiona}
 y^{(N)}(a)=\sum_{n=0}^{N}\epsilon^ny_n(a)=0.
\end{equation}
The condition $ y^{(N)}(a)=0$ identifies the limit cycle $y(x)$ up to the order 
${\mathcal O}(\epsilon^{N})$,
but its amplitude $a$ only up to the order ${\mathcal O}(\epsilon^{N-1})$ because 
$y_0(a)$ vanishes for any value of $a$. Hence, the
last condition (\ref{conditiona}) can be rewritten as
\begin{equation}
\label{condition}
y^{(N)}(a)=0 \hskip 5mm\Longrightarrow\hskip 5 mm\sum_{n=1}^{N}\epsilon^{n-1}y_n(a)=0.
\end{equation}
The solutions $a$ from this last equation are used to calculate
$y_{N}(x)$, and hence the approximation $y^{(N)}(x)$ for the 
form of the limit cycles can be finally obtained. The iteration
of this scheme for $N=1,2,3\ldots$ generates the whole expansion 
of the limit cycle in  powers of $\epsilon$.

\subsection{Application of the method for the lower orders}

\subsubsection{Limit cycles up to the order ${\mathcal O}(\epsilon^1)$ }

Up to the order ${\mathcal O}(1)$ in the amplitude ($N=1$), the above 
condition (\ref{condition}) imposes the
following restriction over the values of $a$:
\begin{equation}
y^{(1)}(a)=0 \hskip 5mm\Longrightarrow\hskip 5mm y_1(a)=0.
\end{equation}
The function $y_1(x)$ reads
\begin{equation}
\label{yone}
 y_1(x)=-{1\over y_0(x)}\int_{-a}^xy_0(t)f(t)dt=-{1\over y_0(x)}\left\lbrace
\beta_0(a)+\int_{a}^xy_0(t)f(t)dt\right\rbrace,
\end{equation}
where $\beta_0(a)$ is given in expression (\ref{beta}).
The function $y_0(t)f(t)$ is continuous in the interval $[-a,a]$ and of the order ${\mathcal O}(\sqrt{a-x})$
when $x\to a$. Therefore,
\begin{equation}
\label{v}
\int_{a}^xy_0(t)f(t)dt={\mathcal O}\left((a-x)^{3/2}\right) \hskip 1cm {\rm when} \hskip 2mm x\to a.
\end{equation}
As $y_0(x)=\sqrt{a^2-x^2}$, then, for any $a>0$,
\begin{equation}
\label{uno}
 y_1(x)=-{\beta_0(a)\over y_0(x)}+{\mathcal O}(a-x),  \hskip 1cm {\rm when} \hskip 2mm x\to a.
\end{equation}
Hence, from (\ref{yone}), a necessary and sufficient condition for $y^{(1)}(x)$ to be zero at
$x=a$ is just the condition (\ref{beta}) derived from the integral equation (\ref{eq4}) 
for the zero order approximation:
\begin{equation}
\beta_0(a)=0.
\label{b00}
\end{equation}
We recover in this way the order ${\mathcal O}(\epsilon^0)$ condition for the amplitudes of the limit cycles
emerging from the period annulus at order zero (\ref{beta}). The form of these solutions is perturbed at the
order ${\mathcal O}(\epsilon^1)$ and reads:
\begin{equation}
y^{(1)}(x)=y_0(x)+\epsilon y_1(x),
\end{equation}
with $y_1(x)$ given by expression (\ref{yonee}) when $a$ is solution of Eq. (\ref{b00}).

\subsubsection{Limit cycles up to the order ${\mathcal O}(\epsilon^2)$ }

Up to order ${\mathcal O}(\epsilon)$ in the amplitude ($N=2$), condition (\ref{condition}) imposes the following
restriction over the values of $a$:
\begin{equation}
\label{yados}
y^{(2)}(a)=0 \hskip 5mm\Longrightarrow\hskip 5mm y_1(a)+\epsilon y_2(a)=0.
\end{equation}
Define
\begin{equation}
\label{betaone}
 \beta_1(a)\equiv\int_{-a}^ay_1(x)f(x)dx.
\end{equation}
Recalling that $f(x)y_0(x)$ is an even function of $x$, and then
$\beta_0(a)=2\int_0^af(x)y_0(x)dx=2\int_{-a}^0f(x)y_0(x)dx$, we have that:
$$
\beta_1(a) =  -\int_{-a}^a{f(x)\over y_0(x)}dx\int_{-a}^xf(t)y_0(t)dt
= -\int_{-a}^a{f(x)\over y_0(x)}dx\left\lbrace{1\over 2}\beta_0(a)+
 \int_{0}^xf(t)y_0(t)dt\right\rbrace.
$$
Because $f(t)y_0(t)$ is an even function of $t$, the $t-$integral in the
above last expression is an odd function of
$x$ and then vanishes with the exterior integral in $x$, when it runs
in the interval $[-a,a]$. The result is
\begin{equation}
\label{buno}
 \beta_1(a)=-{1\over 2}\beta_0(a)\int_{-a}^a{f(x)\over y_0(x)}dx=
 -{1\over 2a}\beta_0(a)\beta_0'(a).
\end{equation}
From this equation and (\ref{ytwo}) we find out that
\begin{equation}
\label{u}
y_2(x)=-{1\over y_0(x)}\left\lbrace{1\over 2}y_1^2(x)
-{1\over 2a}\beta_0(a)\beta_0'(a)+\int_a^x y_1(t)f(t)dt\right\rbrace.
\end{equation}
Observe that if $a$ is a solution of $\beta_0(a)=0$, that is $y_1(a)=0$, the same 
value of $a$ automatically
satisfies $y_2(a)=0$. This means that, up to the order ${\mathcal O}(\epsilon)$, 
the amplitudes of the limit cycles of Eq.
(\ref{eq3}) are the positive roots of the polynomial $\beta_0(a)$. Then, it seems that they do not experience
any correction at the order ${\mathcal O}(\epsilon)$. Let's show this statement more precisely. Write
$a=a_0+a_1\epsilon+{\mathcal O}(\epsilon^2)$ when $\epsilon\to 0$, where $a_0$ is a root of $\beta_0(a)$:
$\beta_0(a_0)=0$ and $a_1$ is a real number. Then, 
\begin{equation}
\label{beo}
\beta_0(a)=\beta_0'(a_0)a_1\epsilon+{\mathcal O}(\epsilon^2).
\end{equation}
In the foregoing discussion we will consider that the limit $\epsilon\to 0$ must be taken
before the limit $x\to a$ and then any product of symbols ${\mathcal O}(\epsilon^p)$ and ${\mathcal
O}\left((a-x)^q\right)$ can be replaced by ${\mathcal O}(\epsilon^p)$. For any $a>0$,
$y_1(x)$ is a continuous functions in $[-a,a]$ up to the order ${\mathcal O}(\epsilon)$ 
and from (\ref{uno}) and (\ref{beo}):
\begin{equation}
\label{dos}
 y_1(x)={-\beta_0(a)\over y_0(x)}+{\mathcal O}(a-x)={\mathcal O}(\epsilon)+{\mathcal O}(a-x).
 \end{equation}
 Then,
\begin{equation}
\int_{a}^xy_1(t)f(t)dt={\mathcal O}\left((a-x)^2\right)+{\mathcal O}(\epsilon)
\hskip 1cm {\rm when} \hskip 2mm x\to a \hskip 5mm {\rm and} \hskip 2mm \epsilon\to 0
\end{equation}
From the above equation and (\ref{u}):
\begin{equation}
\label{yy}
 y_2(x)=-{1\over y_0(x)}\left[{\mathcal O}\left((a-x)^{2}\right)+\beta_1(a)+{\mathcal
O}(\epsilon)\right].
\end{equation}
From (\ref{uno}) and (\ref{yy}) we have that:
$$
y^{(2)}(x)  =  y_0(x)+\epsilon\left[ {\mathcal O}(a-x) -{\beta_0(a)\over y_0(x)}\right]+
\epsilon^2\left[{\mathcal O}\left((a-x)^{3/2}\right)-{\beta_1(a)\over y_0(x)}\right]  +  
{\mathcal O}(\epsilon^3).
$$
Hence, in order to have $y^{(2)}(a)=0$ up to
the order ${\mathcal O}(\epsilon)$, the function
\begin{eqnarray}
\beta^{(1)}(a,\epsilon)\equiv \beta_0(a)+\epsilon\beta_1(a)
\end{eqnarray}
must satisfy
\begin{equation}
\label{beuno} \beta^{(1)}(a,\epsilon)={\mathcal O}(\epsilon^2).
\end{equation}
Using that $\beta_0(a)=\beta_0'(a_0)a_1\epsilon+{\mathcal O}(\epsilon^2)$ and
$\beta_1(a)=\beta_1(a_0)+{\mathcal O}(\epsilon)={\mathcal O}(\epsilon)$ (see (\ref{buno})),
the above equation implies
\begin{equation}
\label{betuno}
\beta_0'(a_0)a_1=0.
\end{equation}
Therefore $a_1=0$ when $\beta_0'(a_0)\neq 0$.
The case $\beta_0'(a_0)= 0$, where a bifurcation of
a multiple limit cycle is possible, is not considered here.
This means that the amplitude of the limit cycles of Eq. (\ref{eq3}) have the form
$a(\epsilon)=a_0+a_2\epsilon^2+{\mathcal O}(\epsilon^3)$, where $a_2$ is a real number.
This is a consequence of the parameter inversion symmetry of Eq. (\ref{eq3})
commented in Section 2.

The form of these solutions is perturbed and reads:
\begin{equation}
\label{y2}
y^{(2)}(x)=y_0(x)+\epsilon y_1(x)+\epsilon^2 y_2(x),
\end{equation}
where $y_1(x),y_2(x)$ are given by expressions (\ref{yonee}-\ref{ytwo}),
with $y_2(x)$ calculated for the solutions $a$ of $\beta_0(a)=0$. 
Observe that as the function $y_0(x)$
is an even function of $x$, then, from (\ref{yone}), we see that $y_1(x)$ is an 
odd function of $x$ when $a$ is
a root of $\beta_0(a)$, and then, from (\ref{ytwo}) and (\ref{buno}), 
it is clear that $y_2(x)$ is an even function of $x$.

\subsubsection{Limit cycles up to the order ${\mathcal O}(\epsilon^3)$ }

Up to order ${\mathcal O}(\epsilon^2)$ in the amplitude ($N=3$), condition 
(\ref{condition}) imposes the
following restriction over the values of $a$:
\begin{equation}
y^{(3)}(a)=0 \hskip 5mm\Longrightarrow\hskip 5mm
y_1(a)+\epsilon y_2(a)+\epsilon^2 y_3(a)=0.
\end{equation}

From the definition (\ref{ythree}) of $y_3(x)$ we have that
\begin{eqnarray}
y_3(x) & = & -{1\over y_0(x)}\left[y_1(x)y_2(x)+\int_{-a}^x y_2(t)f(t)dt\right]= \\
& = & -{1\over y_0(x)}\left[y_1(x)y_2(x)+\int_a^x y_2(t)f(t)dt+\int_{-a}^a y_2(t)f(t)dt\right].
\end{eqnarray}

From the previous section we have that $\beta_0(a)=\beta_0'(a_0)a_2\epsilon^2+{\mathcal O}(\epsilon^3)$ and
$\beta_1(a)={\mathcal O}(\epsilon^2)$. Then, from (\ref{uno}) and (\ref{u}) we have that, 
for any $a>0$ and up to the order ${\mathcal O}(\epsilon^2)$,
$y_1(x)$ and $y_2(x)$ are continuous functions in $[-a,a]$, 
$y_1(x)={\mathcal O}(a-x)+{\mathcal O}(\epsilon^2)$ and
\begin{equation}
\label{tres}
 y_2(x)={\mathcal O} \left((a-x)^{3/2}\right)+{\mathcal O}(\epsilon^2).
 \end{equation}
 Then:
\begin{equation}
\label{z}
\int_{a}^xy_2(t)f(t)dt={\mathcal O}\left((a-x)^{5/2}\right)+{\mathcal O}(\epsilon^2)
\hskip 1cm {\rm when} \hskip 2mm x\to a \hskip 5mm {\rm and} \hskip 2mm \epsilon\to 0
\end{equation}
and
\begin{equation}
\label{cuatro}
 y_3(x)=-{1\over y_0(x)}\left[{\mathcal O}\left((a-x)^{5/2}\right)+\beta_2(a)+{\mathcal
O}(\epsilon^2)\right],
\end{equation}
where we have defined
\begin{equation}
\beta_2(a)\equiv \int_{-a}^a y_2(t)f(t)dt.
\end{equation}
If the results (\ref{uno}), (\ref{tres}) and (\ref{cuatro}) are substituted in the expression (\ref{y3}) we
obtain:
\begin{eqnarray}
y^{(3)}(x) & = & y_0(x)+\epsilon\left[ {\mathcal O}(a-x) -{\beta_0(a)\over y_0(x)}\right]+
\epsilon^2\left[ {\mathcal O}\left((a-x)^{3/2}\right)+ {\mathcal O}(\epsilon^2)\right]+  \nonumber\\
& & +\epsilon^3\left[{\mathcal O}\left((a-x)^2\right) -{\beta_2(a)\over y_0(x)}+
{\mathcal O}(\epsilon^2)\right]+ {\mathcal O}(\epsilon^4).
\end{eqnarray}
Using that $\beta_0(a)=\beta_0'(a_0)a_2\epsilon^2+
{\mathcal O}(\epsilon^3)$ and  $\beta_2(a)=\beta_2(a_0)+{\mathcal O}(\epsilon)$ we have that
\begin{eqnarray}
y^{(3)}(x) & = & y_0(x)+\epsilon\left[ {\mathcal O}(a-x) -{\beta_0'(a_0)a_2\over y_0(x)}\epsilon^2\right]+
\epsilon^2{\mathcal O}\left((a-x)^{3/2}\right)+  \nonumber\\
& & +\epsilon^3\left[{\mathcal O}\left((a-x)^{2}\right) -{\beta_2(a_0)\over y_0(x)}\right]+
{\mathcal O}(\epsilon^4).
\end{eqnarray}

Hence, in order to have $y^{(3)}(a)=0$ up to
the order ${\mathcal O}(\epsilon^2)$, the function
$$
\beta^{(2)}(a,\epsilon)\equiv \beta_0(a)+\epsilon\beta_1(a)+\epsilon^2\beta_2(a)
=\left[\beta_0'(a_0)a_2+\beta_2(a_0)\right]\epsilon^2+ {\mathcal O}(\epsilon^3)
$$
must satisfy
\begin{equation}
\label{be} \beta^{(2)}(a,\epsilon)={\mathcal O}(\epsilon^3).
\end{equation}
Therefore
\begin{equation}
a_2=-{\beta_2(a_0)\over \beta_0'(a_0)}.
\end{equation}
The correction of the amplitude up to order ${\mathcal O}(\epsilon^2)$
is finally obtained:
\begin{equation}
a(\epsilon)=a_0-{\beta_2(a_0)\over \beta_0'(a_0)}\;\epsilon^2+{\mathcal O}(\epsilon^3),
\label{eqbe1}
\end{equation}
and the form of the solutions at the order ${\mathcal O}(\epsilon^3)$ reads:
\begin{equation}
y^{(3)}(x)=y_0(x)+\epsilon y_1(x)+\epsilon^2 y_2(x)+\epsilon^3 y_3(x),
\end{equation}
where $y_1(x),y_2(x),y_3(x)$ are given by expressions (\ref{yonee}-\ref{ythree}),
with $y_3(x)$ calculated for the amplitudes $a(\epsilon)$ given by the 
expression (\ref{eqbe1}).

\section{An alternative view: \\ the recursion of the integral equation}

Let us see that the method explained in the last section can be viewed in
an equivalent and alternative way as a recursive approximation
in successive powers of $\epsilon$ to the integral Eq. (\ref{eq4}).

\subsection{General algorithm}

\noindent\underline{Step 0}: Set $y_0(x)=\sqrt{a^2-x^2}$ and $N=0$.

\noindent\underline{Step 1}: Define the approximation of $y(x)$ at order 
${\mathcal O}(\epsilon^N)$ given by Eq.
(\ref{eqyn}):
\begin{equation}
y(x)\simeq y^{(N)}(x)\equiv \sum_{n=0}^{N} \epsilon^ny_n(x),
 \label{eqyn1}
\end{equation}
and find the solutions $a$ of the integral equation at order ${\mathcal O}(\epsilon^N)$:
\begin{equation}
\label{betan}
\beta^{(N)}(a,\epsilon)\equiv\int_{-a}^af(x)y^{(N)}(x)dx=0,
\end{equation}
that is equivalent to the equation
\begin{equation}
\beta^{(N)}(a,\epsilon)\equiv\sum_{n=0}^{N} \epsilon^n\beta_n(a)=0,
\label{eqyn11}
\end{equation}
where
\begin{equation}
\label{betann}
\beta_n(a)\equiv\int_{-a}^af(x)y_n(x)dx=0.
\end{equation}

\noindent\underline{Step 2}: For each solution $a$ of ${\it step\hskip 2mm 1}$ 
calculate $y_{N+1}(x)$ by the
formula (\ref{eqw}):
\begin{eqnarray}
y_{N+1}(x) & = & -{1\over y_0(x)}\int_{-a}^xW_{N}(t)dt = \nonumber \\
 & = &  -{1\over y_0(x)}\left\lbrace {1\over
2}\sum_{k=1}^{N}y_k(x)y_{n-k}(x)+\int_{-a}^xy_{N}(t)f(t)dt\right\rbrace.
\label{eqw1}
\end{eqnarray}

\noindent\underline{Step 3}: Replace $N$ by $N+1$ and come back to {\it step 1}.

{\bf Note:} It can be easily found that the application of this algorithm for 
the lower orders, up to $N=2$,
repeats the results obtained in Section 3.3.

\section{Examples}

The limit cycles in the weak and in the strongly nonlinear regimes
of different families of Li\'enard systems were studied in \cite{lopezruiz1}.
Here we perform the calculations proposed in Section 3 for two concrete examples,
which are particular cases of the families $1$ and $3$ worked out in \cite{lopezruiz1}.

{\bf Example 1.} The van der Pol oscillator is given for $f(x)=x^2-1$.
This system has a unique limit cycle, which is stable for $\epsilon>0$.
Hence, the only root of $\beta_0(a)$ is $a_0=2$. For this value of $a_0$, we have
$y^{(2)}(x)=y_0(x)+\epsilon y_1(x)+\epsilon^2 y_2(x)$ with
$$
y_0(x)=\sqrt{4-x^2},
$$
$$
y_1(x)={x\over 4}(4-x^2)
$$
and
$$
y_2(x)={2+x^2\over 96}(4-x^2)^{3/2}.
$$

We integrate Eq. (\ref{eq3}) by a Runge-Kutta method in order
to obtain the limit cycle. This curve is plotted in a continuous trace in Fig. 1(a-b)
for $\epsilon=0.5$ and $\epsilon=1$, respectively. The approximated limit
cycle $y^{(2)}(x)$ is also plotted in those figures with a discontinuous trace
for the same values of $\epsilon$.
Let us remark that, in this case, even up to $\epsilon=3$,
the approximation $y^{(2)}(x)$ to the limit cycle is very good.

The solution of $\beta^{(2)}(a,\epsilon)=0$, up to order ${\mathcal O}(\epsilon^2)$,
gives us:
$$
\beta_2(a_0)=-{\pi\over 48},\hskip 1cm  \beta_0(a)={\pi\over 8}a^2(a^2-4),
$$
then
\begin{equation}
a(\epsilon)=2+{1\over 96}\;\epsilon^2+{\mathcal O}(\epsilon^3).
\label{aa}
\end{equation}
We must stress at this point that this expansion was also done for the van der Pol
system in \cite{vanhorssen} by a perturbation method based on integrating factors.
The author reported in that article \cite{vanhorssen} the expansion:
\begin{equation}
a(\epsilon)=2+{23\over 96}\;\epsilon^2+{\mathcal O}(\epsilon^3).
\label{aaa}
\end{equation}
Our calculation (\ref{aa}) does not agree with this result, but agrees
with the computational calculation of the 'exact' amplitudes given in Table 1.

{\bf Example 2.} The same process is performed for $f(x)=5x^4-9x^2+1$.
In this case, the system has two limit cycles, one stable and the other unstable.
The polynomial $\beta_0(a)$ has two positive roots:
$a_0=\sqrt{9-\sqrt{41}\over 5}=0.720677$ (unstable limit cycle for $\epsilon>0$)
and $\bar a_0=\sqrt{9+\sqrt{41}\over 5}=1.755170$ (stable limit cycle for $\epsilon>0$).

For the first limit cycle we have $y^{(2)}(x)=y_0(x)+\epsilon y_1(x)+\epsilon^2 y_2(x)$ with
$$
y_0(x)=\sqrt{a_0^2-x^2},
$$
$$
y_1(x)={x\over 24}(27+3\sqrt{41}-20x^2)(x^2-a_0^2)
$$
and
$$
y_2(x)={(a_0^2-x^2)^{3/2}\over 2880}\left[{3222-218\sqrt{41}
\over 5}+(1003+63\sqrt{41})x^2-15(27+7\sqrt{41})x^4+200x^6\right].
$$

The solution of $\beta^{(2)}(a,\epsilon)=0$, up to order ${\mathcal O}(\epsilon^2)$,
gives us:
$$
\beta_2(a_0)=-0.007357,\hskip 1cm  \beta_0(a)={\pi\over 16}a^2(5a^4-18a^2+8).
$$
Then
\begin{equation}
\label{ab}
a(\epsilon)=0.720677+0.003908\;\epsilon^2+{\mathcal O}(\epsilon^3).
\end{equation}

For the second limit cycle we have $y^{(2)}(x)=y_0(x)+\epsilon y_1(x)+\epsilon^2 y_2(x)$ with
$$
y_0(x)=\sqrt{\bar a_0^2-x^2},
$$
$$
y_1(x)=x\left({9-\sqrt{41}\over 8}-{5\over 6}x^2\right)(x^2-\bar a_0^2)
$$
and
$$
y_2(x)={(\bar a_0^2-x^2)^{3/2}\over 2880}\left[{3222+218\sqrt{41}
\over 5}+(1003-63\sqrt{41})x^2+15(7\sqrt{41}-27)x^4+200x^6\right].
$$

The solution of $\beta^{(2)}(a,\epsilon)=0$, up to order ${\mathcal O}(\epsilon^2)$,
gives us:
$$
\beta_2(\bar a_0)=-0.486199,\hskip 1cm  \beta_0(a)={\pi\over 16}a^2(5a^4-18a^2+8).
$$
Then
\begin{equation}
\label{ac}
a(\epsilon)=1.755170-0.017880\;\epsilon^2+{\mathcal O}(\epsilon^3).
\end{equation}

The comparison between the 'exact' and the approximated limit cycles
can be seen in the plots of Fig. 2(a-b) and Fig. 3(a-b).

\section{Conclusions}

A general algorithm to approximate the form and the amplitude
of limit cycles in the weakly nonlinear regime of Li\'enard equations
has been presented. In each iteration of the method a new order
in $\epsilon$ is calculated. Thus, the new term added in the expansion
of $\epsilon$ for the form of the limit cycle must verify the differential
equation, and the correction to this order for the amplitude of the limit
cycle is derived from the integral equation. Different examples have been
worked out and the results are in good agreement with the direct integration
of the limit cycles.

In this paper we only have detailed the approximation up to the order $\epsilon^3$.
This process can be rewritten in a formalism that allow us to go
farther in the order of approximation. We are just working in this
direction and these new results will be presented elsewhere.

{\bf Acknowledgements:}  R. L-R. acknowledges financial support from the spanish 
research project FIS2004-05073-C04-01. 
J. L. L. acknowledges the financial support of the {\it Direcci\'on General de Ciencia y
Tecnologia (REF. MTM2004-05221)}(Spain).

%\newpage

\newpage

\noindent{\bf Table 1.} The value $a_{T}$ represents the approximated
amplitude $a(\epsilon)$ of the van der Pol limit cycle
obtained from (\ref{aa}) for the indicated values of $\epsilon$.
The value $a_{E}$ represents the amplitude $a$ obtained by integrating directly
the system with a Runge-Kutta method.

\begin{center}
\begin{normalsize}
\begin{flushleft}
\begin{tabular}{|p{.6cm}|p{2cm}|p{2cm}|p{2cm}|p{2cm}|l|l|}
\hline $\epsilon$ & $0.1$ & $0.2$ & $0.3$ & $0.4$ & $0.5$ \\
\hline $a_{T}$&2.00010&2.00041&2.00093&2.00166&2.00260\\
\hline $a_{E}$&2.00010&2.00041&2.00092&2.00161&2.00248\\
\hline
\hline $\epsilon$ & $0.6$ & $0.7$ & $0.8$ & $0.9$ & $1$\\
\hline $a_{T}$&2.00375&2.00510&2.00666&2.00843&2.01041\\
\hline $a_{E}$&2.00351&2.00466&2.00591&2.00724&2.00862\\
\hline
\end{tabular}
\end{flushleft}
\end{normalsize}
\end{center}

\begin{center} {\bf Figures} \end{center}

{\bf Fig. 1a-b}:
Exact limit cycle (continuous line) and approximated limit cycle (discontinuous line )
up to order ${\mathcal O}(\epsilon^2)$ for the van der Pol system with
(a) $\epsilon=0.5$ and (b) $\epsilon=1$.

{\bf Fig. 2a-b}:
Exact limit cycle (continuous line) and
approximated limit cycle (discontinuous line ) up to order
${\mathcal O}(\epsilon^2)$ for the smallest limit cycle
of the example 2 with (a) $\epsilon=-0.5$ and (b) $\epsilon=-1$.

{\bf Fig. 3a-b}:
Exact limit cycle (continuous line) and
approximated limit cycle (discontinuous line ) up to order
${\mathcal O}(\epsilon^2)$ for the biggest limit cycle
of the example 2 with (a) $\epsilon=0.5$ and (b) $\epsilon=1$.

\newpage

\begin{figure}[]
\includegraphics[angle=0, width=15cm]{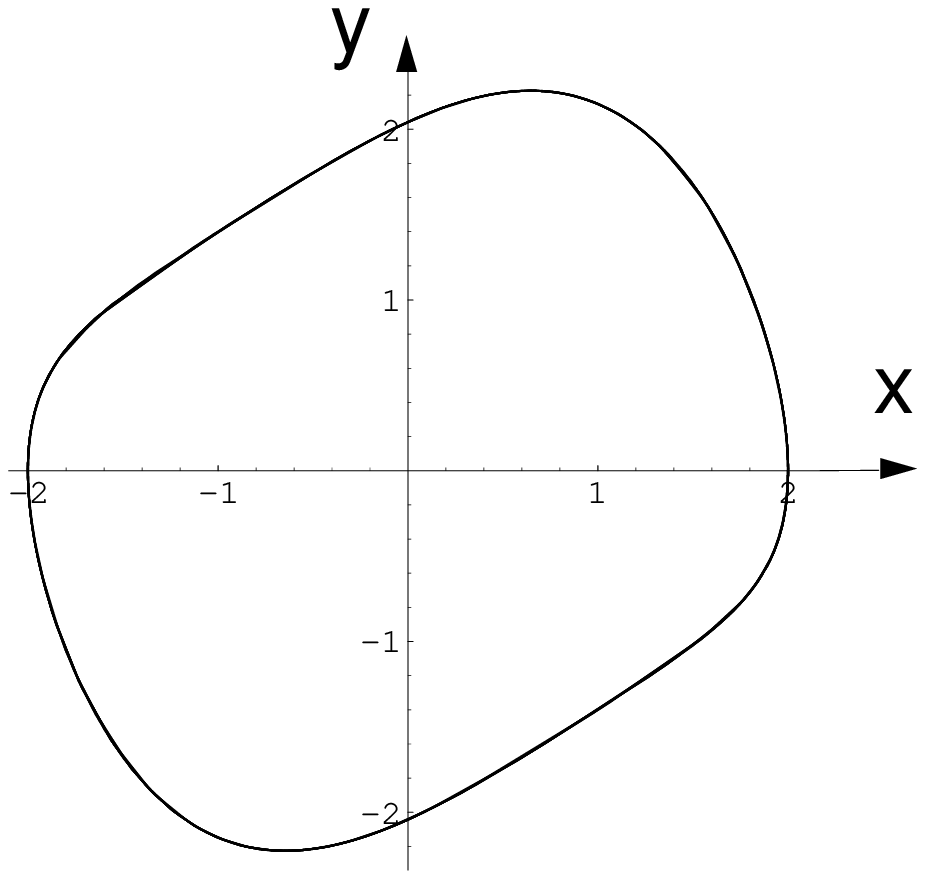}
\label{fig1a}
\end{figure}

\begin{figure}[]
\includegraphics[angle=0, width=15cm]{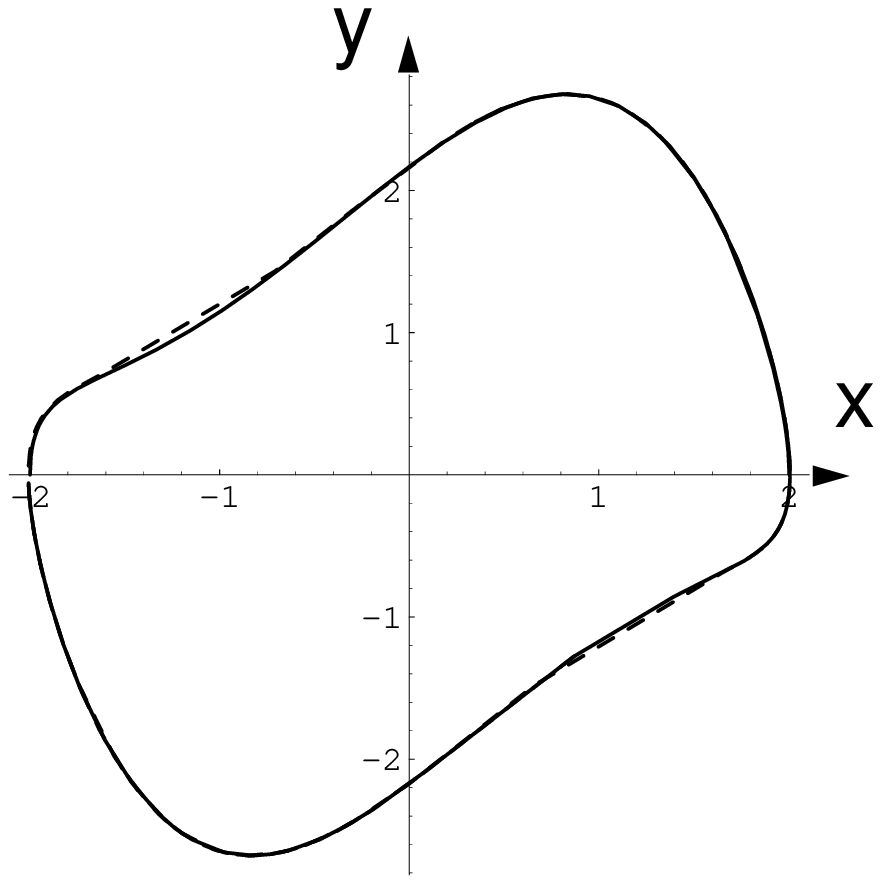}
\label{fig1b}
\end{figure}

\newpage
\begin{figure}[]
\includegraphics[angle=0, width=15cm]{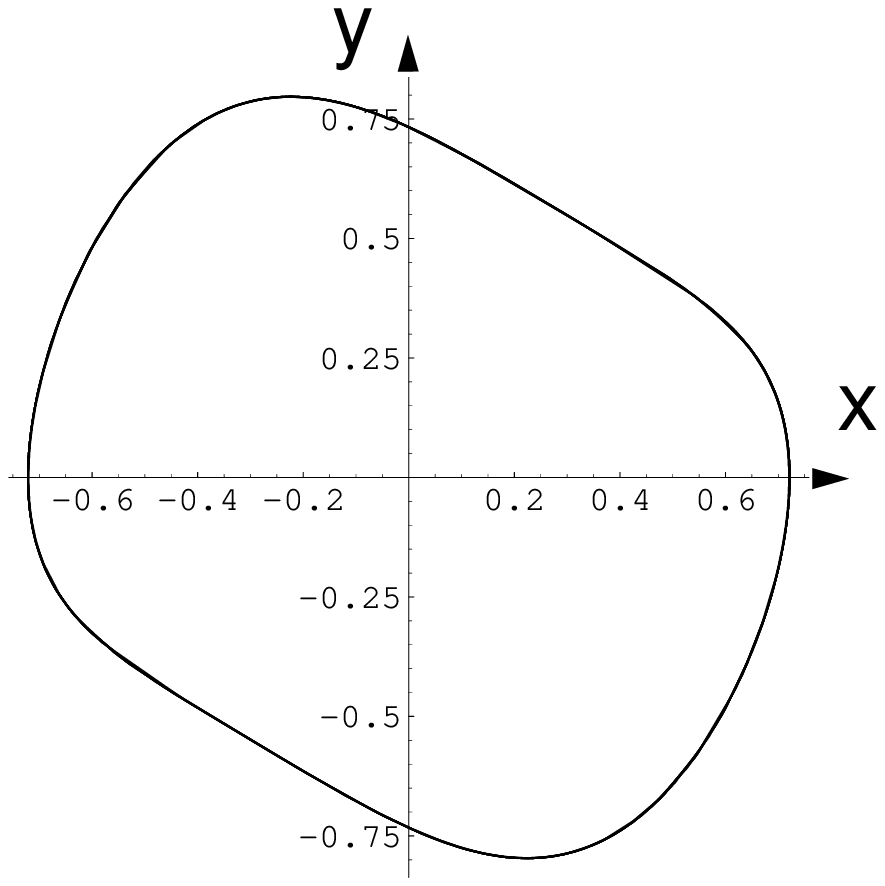}
\label{fig2a}
\end{figure}

\begin{figure}[]
\includegraphics[angle=0, width=15cm]{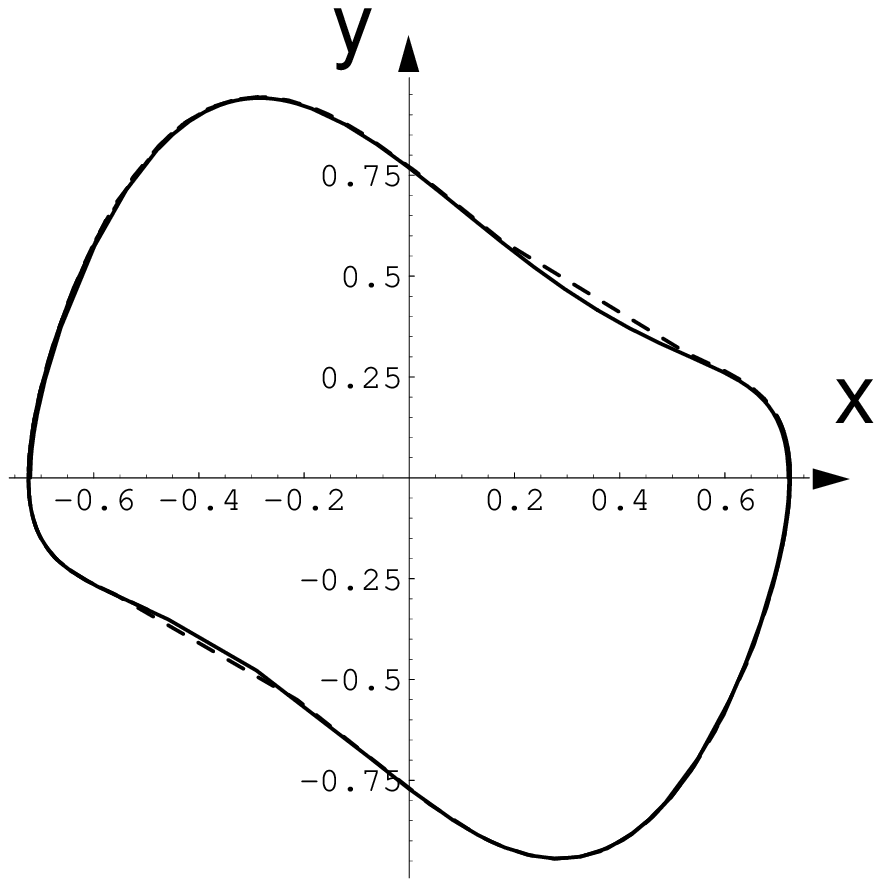}
\label{fig2b}
\end{figure}

\begin{figure}[]
\includegraphics[angle=0, width=15cm]{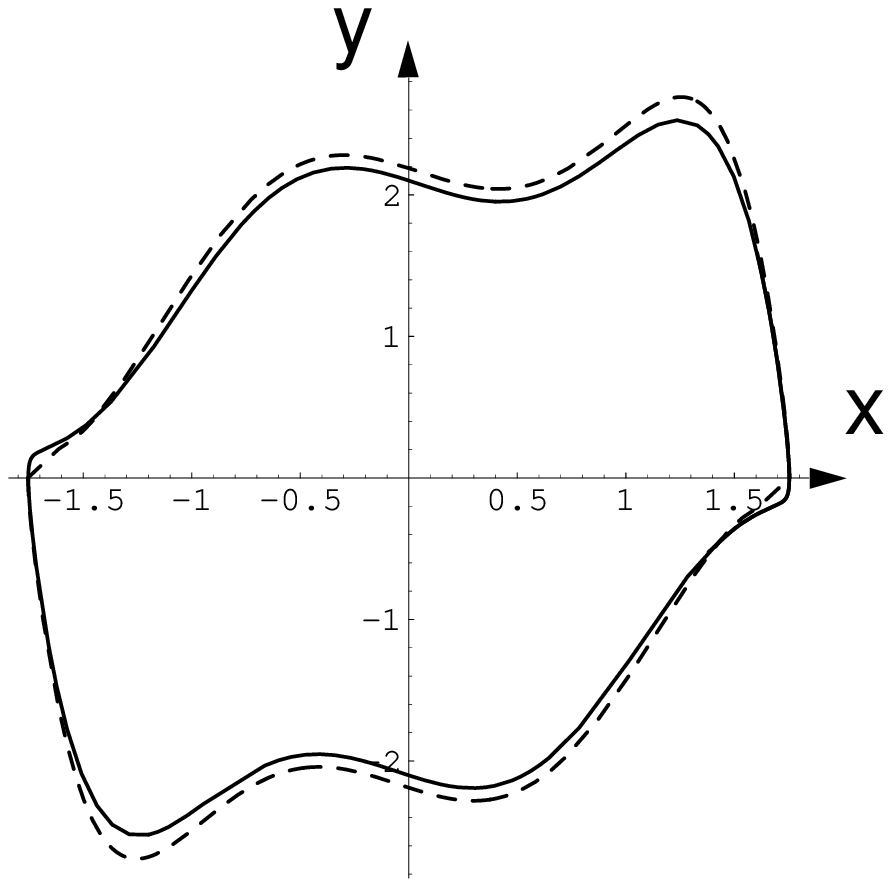}
\label{fig3a}
\end{figure}

\begin{figure}[]
\includegraphics[angle=0, width=15cm]{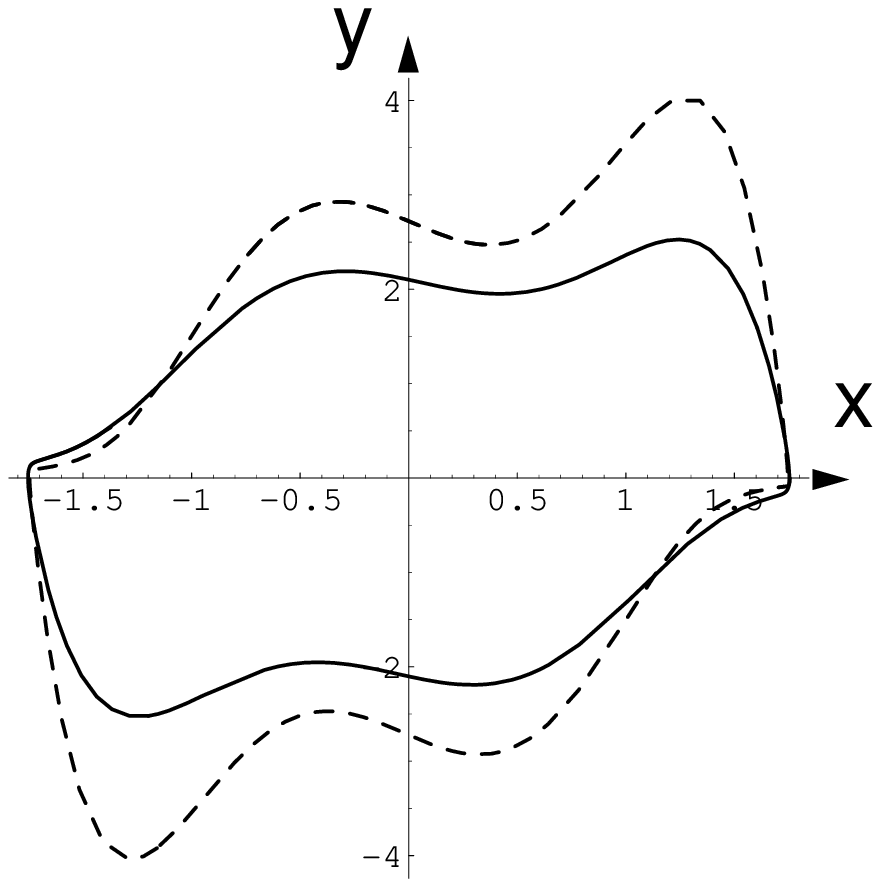}
\label{fig3b}
\end{figure}

\end{document}